\documentclass[%
reprint,
 amsmath,amssymb,
 aps,
 pra,
]{revtex4-2}

\usepackage[
CJKbookmarks=false,
colorlinks,
breaklinks=true,
linkcolor=blue,
anchorcolor=blue,
citecolor=blue,
urlcolor=blue,
]{hyperref}

\usepackage{graphicx}
\usepackage{dcolumn}
\usepackage{bm}

\begin{document}

\preprint{APS/123-QED}

\title{Efficient characterization of quantum nondemolition qubit readout}

\author{He Wang}%
 \email{wanghe22@baidu.com}
\author{Ya Cao}
\email{caoya@baidu.com}
\affiliation{%
Institute for Quantum Computing, Baidu Research, Beijing 100193, China
}%

\date{\today}

\begin{abstract}
We study the quantitative characterization of the performance of qubit measurements in this paper. In particular, the back-action evading nature of quantum nondemolition (QND) readout of qubits is fully quantified by quantum trace distance. Only computational basis states are necessary to be taken into consideration. Most importantly, we propose an experimentally efficient method to evaluate the QND fidelity based on the classical trace distance, which uses an experimental scheme with two consecutive measurements. The three key quantifiers of a measurement, i.e., QND fidelity, readout fidelity, and projectivity, can be derived directly from the same experimental scheme. Besides, we present the relationships among these three factors. Theoretical simulation results for the dispersive readout of superconducting qubits show the validity of the proposed QND fidelity. Efficient quantification of measurement performance is of practical significance for the diagnosis, performance improvement, and design of measurement apparatuses.
\end{abstract}


\maketitle

\section{Introduction}
Quantum measurement links the classical and quantum worlds \cite{wiseman2009quantum} and plays a vital role in quantum computation and quantum information \cite{nielsen2010quantum}. However, quantum measurements are vulnerable to noises \cite{clerk2010introduction}. Characterizing and improving the performance of quantum measurement is essential. Quantum non-demolition (QND) measurement preserves the expectation value of an observable, and thus enables repetitive measurement. The repeatability of QND measurements is of great significance in improving measurement performance \cite{Vladimir1980Quantum,Caves1980On,Bocko1996On,Braginsky1996Quantum,Hume2007High,grangier1998quantum,orenes2022improving}. In quantum computing, one of the key goals is to develop a high-performance measurement of qubits. Qubit readout is generally required to meet the QND requirement \cite{lupacscu2007quantum,Gambetta2007Protocols,nakajima2019quantum,yoneda2020quantum,raha2020optical,gusenkova2021quantum,Dassonneville2020Fast}. Besides, the QND nature of a measurement is essential in repetitive quantum error correction \cite{schindler2011experimental,kelly2015state,Xue2020Repetitive,krinner2022realizing,Zhao2022Realization,marques2022logical,Rajeev2022Suppressing}, reducing measurement-induced state mixing error \cite{Slichter2012Measurement}, and heralded state preparation \cite{Johnson2012Heralded}.

Nowadays, it's still a challenge to improve the QND performance in qubit readout. In superconducting quantum computing, the dispersive readout is the standard approach to qubit readout and plays a central role in most circuit quantum electrodynamics (QED) experiments \cite{krantz2019quantum,kwon2021gate,Blais2021Circuit,wallraff2004strong}. However, the dispersive readout is highly constrained and will lose the QND nature when the dispersive limit is broken \cite{Slichter2012Measurement,Blais2021Circuit}. To enhance the QND character of quantum measurements, a variety of quantum measurement schemes have been developed \cite{Didier2015Fast,Touzard2019Gated,Dassonneville2020Fast}.

Theoretical and experimental quantifications of the QND property are essential in the diagnosis and improvement of a measurement apparatus \cite{pereira2021complete,pereira2022parallel}. The commonly used experimental scheme for the characterization of the QND-ness of a measurement is constructed of two consecutive measurements \cite{lupacscu2007quantum}, and the QND-ness is quantified by the probability of obtaining the same outcomes from these two measurements \cite{lupacscu2007quantum,Picot2010Quantum,Touzard2019Gated,Dassonneville2020Fast}. However, the QND-ness defined in this way actually qualifies its \emph{projectivity} or \emph{ideality}, i.e., the overlap with an ideal projective measurement, rather than the QND character of the measurement \cite{pereira2021complete}. It should be noted that non-ideal QND measurements are important in quantum information tasks that require precise evaluations of expectation values \cite{pereira2021complete}, such as variational quantum eigensolver algorithms \cite{peruzzo2014variational,McArdle2020Quantum,cerezo2021variational} and quantum sensing \cite{Degen2017Quantum}.

To characterize the physical nature of a QND measurement, L. Pereira \emph{et al.} have introduced a self-consistent tomography for detectors, which consists of two consecutive applications of a detector, in between an unitary operation from a universal set of gates is interleaved \cite{pereira2021complete}. There they apply quantum process tomography to reconstruct all the Choi matrices of the measurement process and accordingly provide a quantification of the measurement destructiveness by solving a series of optimization problems. However, the computational complexity of this method is very high since the high complexity of process tomography. Besides, a universal set of quantum states and gates are utilized to perform quantum tomography, which is challenged by state preparation and quantum gate errors.

In this paper, we propose a novel method to efficiently characterize the back-action evading nature of a QND measurement. Only computational basis states are needed. The quantum trace distance between density matrices of the quantum states before and after a measurement is utilized to quantify the back-action of the measurement and thus be used for defining the theoretical QND fidelity. To realize an experimentally efficient characterization of the physical nature of a QND measurement, the measurement is applied twice consecutively. The probability distributions of outcomes from the first and second measurements are compared to evaluate the measurement \emph{demolition}, i.e., the back-action of the measurement. The experimental QND fidelity can be derived directly using the demolition and provides a good estimate for the theoretical QND fidelity. In this way, we greatly lower down the complexity of quantifying the QND nature. Additionally, the experimental QND fidelity and readout fidelity together are sufficient to characterize the projectivity of a measurement. We also present the relationships among key factors of a measurement, i.e., readout fidelity, projectivity, and QND fidelity. Besides, these key factors can be derived directly with the same scheme by using different postprocessing methods. Our work is quite practical in improving the performance of quantum measurements and the design of measurement devices.

The structure of this article is as follows. In Sec.~\ref{sec:notation}, we introduce basic notations and clarify the definition of a QND measurement and several equivalent formations. In Sec.~\ref{sec:QNDdefinition}, we demonstrate the proposed precise quantifier of a measurement, i.e., the demolition and theoretical QND fidelity. In Sec.~\ref{sec:ExperimentalMethod}, we propose an experimental method and obtain the experimental QND fidelity, which is a reachable upper bound for the theoretical QND fidelity. In Sec.~\ref{sec:relationship}, we analyze the relationships among readout fidelity, projectivity, theoretical QND fidelity, and experimental QND fidelity we proposed. Sec.~\ref{sec:simulation} shows the theoretical simulation results for the dispersive readout of superconducting qubits. Sec.~\ref{sec:conclusion} contains the conclusions.

\section{quantum nondemolition qubit readout}\label{sec:notation}
Suppose an observable $O$ which is generally a Hermitian operator satisfies
\begin{equation}\label{eq:QNDHamiltonian}
  [H, O] = 0,
\end{equation}
where $H$ is the Hamiltonian of the composite system consisting of the quantum system to be measured and the assisted measurement system.
The observable $O$ stays unchanged in the Heisenberg picture, i.e., $dO/dt = 0$. Then the observable $O$ is called a QND observable \cite{Vladimir1980Quantum,Caves1980On,Bocko1996On}. 

In a QND measurement, for any density operator $\rho$, the expectation value $\langle O\rangle = \mathrm{Tr}[O\rho]$ satisfies
\begin{equation}
\frac{d\langle O\rangle}{dt} = 0,
\end{equation}
which shows that a QND measurement preserves the expectation value of the observable, called the back-action evading nature of a QND measurement.

In qubit readout, measurements are performed traditionally on the computation basis $\{|k\rangle\langle k |\}$ with $k=0,1,2,...,N$, where $N$ is the number of projection operators and also is the dimension of the Hilbert space. We have $\sum_k |k\rangle\langle k | = I$.

A quantum measurement can be described by Kraus operators $\{M_m\}$ \cite{nielsen2010quantum}. Suppose an arbitrary density operator $\rho$, the unnormalized state conditioned on the outcome $m$ should be 
\begin{equation}
\mathcal{E}_m(\rho) = M_m \rho M_m^{\dagger},
\end{equation}
with probability $p_m(\rho) = \mathrm{Tr}[\mathcal{E}_m(\rho)]$. Notice that the measurement has $N$ outcomes. It stands that $\sum_m M_m^{\dagger}M_m = I$.

Denote the non-selective state after the measurement as $\mathcal{E}(\rho) = \sum_m\mathcal{E}_m(\rho)$, and the corresponding POVM element as $E_m = M_m^{\dagger}M_m$, we have that
\begin{equation}
  p_m(\rho) = \mathrm{Tr}[E_m\rho],
  \end{equation}
where $\sum_m E_m = I$.

According to the fact that the expectation value of an observable $O$ keeps unchanged in a QND measurement, we can describe the QND measurement by using the formalism of quantum operations. By rewriting $dO/dt = 0$, we have that \cite{wiseman2009quantum}
\begin{equation}
  O \otimes I_{A} = U^{\dagger} (O \otimes I_{A}) U,
\end{equation}
where the unitary operator $U=\exp(-iHt)$ is the evolution operator of the composite system. We set $\hbar = 1$ throughout the text with $\hbar$ being the reduced Planck constant. The assisted measurement system is symboled as $A$. Denote $I_A$ to be the identity operator on system $A$, and $\rho\otimes\sigma_{A}$ to be the initial state on the composite system. By multiplying $\rho\otimes\sigma_{A}$ on both sides of the above equation and taking a total trace, one obtains the following,
\begin{equation}\label{eq:QNDobservable}
  \langle O \rangle = \mathrm{Tr}[O\rho] = \mathrm{Tr}[O\mathcal{E}(\rho)],
\end{equation}
where $\mathcal{E}(\rho) = \mathrm{Tr}_{A}[U (\rho\otimes\sigma_{A}) U^{\dagger}]$ is the non-selective state after the measurement. Based on this, one can define a QND qubit readout, i.e., any diagonal observable $O$ in computational basis satisfies Eq.~(\ref{eq:QNDobservable}) for any density operator $\rho$. The definition can be simplified as 
\begin{equation}\label{eq:MeasurementInducedDephasing}
  \langle k |\rho |k\rangle = \langle k |\mathcal{E}(\rho) |k\rangle,~\forall k.
\end{equation}
Physically, the definition above shows that the non-selective state after a QND measurement keeps its diagonal elements unchanged for any input state with density matrix $\rho$. This implies that a QND measurement would not induce state transitions \cite{Slichter2012Measurement}, which is important for reducing readout errors. Based on Eq.~(\ref{eq:QNDobservable}), for arbitrary density matrix $\rho$, one obtains 
\begin{equation}
  \mathrm{Tr}[O\rho] = \mathrm{Tr}[\mathcal{E}^{\dagger}(O)\rho],
\end{equation}
where $\mathcal{E}^{\dagger}(O) = \sum_m M_m^{\dagger} O M_m$. One can simplify it as 
\begin{equation}\label{eq:DestructivenessDefination}
  \mathcal{E}^{\dagger}(O) = O.
\end{equation}
By the fact that $O$ is an arbitrary diagonal observable, the definition above is also equivalent to the following,
\begin{equation}\label{eq:QNDAdjoint}
  \mathcal{E}^{\dagger}(|k\rangle\langle k |) = |k\rangle\langle k |,~\forall k.
\end{equation}
In appendix \ref{appendix:QNDKrausQNDAdjoint}, we have proven that Eq.~(\ref{eq:QNDAdjoint}) is equivalent to that the Kraus operators $\{M_n\}$ are all diagonal, i.e.,
\begin{equation}\label{eq:QNDKraus}
  M_n = \sum_k m_{n,k} |k\rangle\langle k |,~\forall n,
\end{equation}
where $m_{n,k} = \langle k | M_n |k\rangle$. Also in appendix \ref{appendix:QNDKrausQNDAdjoint}, we have claimed that Eq.~(\ref{eq:QNDKraus}) is equivalent to
\begin{equation}\label{eq:QNDKraus2}
  \mathcal{E}(|k\rangle\langle k |) = |k\rangle\langle k |,~\forall k.
\end{equation}
This illustrates that, for a QND measurement, when the computational basis state is measured, the non-selective state after the measurement keeps unchanged. This is the core property to define the measurement demolition and QND fidelity in the latter of this paper.

If the qubit is considered as an ideal two-level system, the requirement for QND qubit readout is $[H, \sigma_z] = 0$, i.e., $[M_m, \sigma_z] = 0$, which is also equivalent to $\mathcal{E}(|k\rangle\langle k |) = |k\rangle\langle k |,~k=0,1$.

\section{Measurement Demolition and QND fidelity}\label{sec:QNDdefinition}
In this section, we define the measurement demolition by quantum trace distance, and then the theoretical QND fidelity. In Sec.~\ref{sec:notation}, we have demonstrated an sufficient and necessary condition Eq.~(\ref{eq:QNDKraus2}) for a QND measurement. Naturally, we have another sufficient and necessary condition
\begin{equation}
  D(|k\rangle\langle k |, \mathcal{E}(|k\rangle\langle k |)) = 0,~\forall k,
\end{equation} 
where $D(\rho,\sigma)$ represents the quantum trace distance between the density matrices $\rho$ and $\sigma$. 
Then we can define the theoretical measurement demolition as follows,
\begin{equation}\label{eq:TraceDistance}
  D_{D}= \frac{1}{N}\sum_k D(|k\rangle\langle k |, \mathcal{E}(|k\rangle\langle k |)).
\end{equation}
We can define the theoretical QND fidelity further as
\begin{equation}
  Q_{D} = 1-D_{D}.
\end{equation}
We can conclude that a measurement is a QND measurement if and only if $Q_{D}=1$. The theoretical QND fidelity $Q_{D}$ is fully determined by the initial state $|k\rangle\langle k |$ and the non-selective state $\mathcal{E}(|k\rangle\langle k |)$ after the measurement. One only needs to take computation basis states into account for the characterization of a QND measurement.

\section{Experimental method}\label{sec:ExperimentalMethod}

\subsection{Experimental setup}
\begin{figure}[tb]
  \centering
  \includegraphics[width=0.48\textwidth]{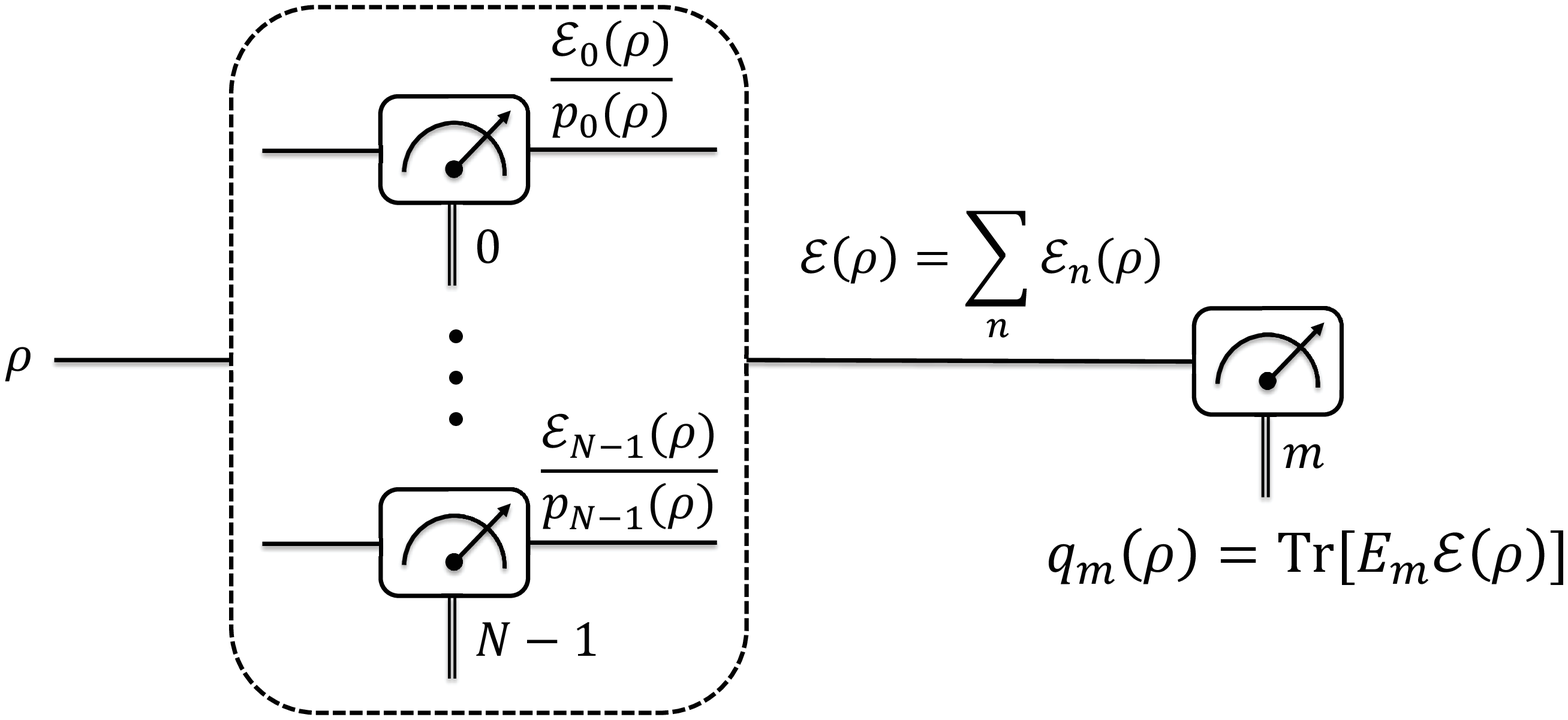}
  \caption{Schematic diagram of the experimental scheme to obtain experimental QND fidelity. For density matrix $\rho$, the probability of the first measurement outputting $m$ is $p_m(\rho)$, while the probability of the second measurement outputting $m$ is $q_m(\rho)$, which relates to the non-selective state $\mathcal{E}(\rho)$ after the first measurement. In our proposed method, only computational basis states are needed to be prepared to obtain the QND fidelity.}
  \label{fig:method2}
  \end{figure}

In order to experimentally and efficiently evaluate the QND character of a measurement, we use an experimental scheme with two consecutive applications of the measurement, as shown in Fig.~\ref{fig:method2}. For an arbitrary density matrix $\rho$, we denote the probability for the first measurement outcome $m$ as $p_m(\rho)$, and the probability for the second measurement outcome $m$ as $q_m(\rho)$. We have $p_m(\rho) = \mathrm{Tr}[E_m \rho]$. Denote the probability for the second measurement outcome $m$ conditioned on the first measurement outcome $n$ as $p_{m|n}(\rho)$, then it stands
\begin{equation}
  p_{m|n}(\rho) = \frac{\mathrm{Tr}[E_m \mathcal{E}_n(\rho)]}{p_n(\rho)}.
\end{equation}
Thus we have
\begin{equation}
  q_{m}(\rho) = \sum_n p_n(\rho)p_{m|n}(\rho) = \mathrm{Tr}[E_m \mathcal{E}(\rho)] = p_{m}(\mathcal{E}(\rho)).
\end{equation}
Notice that the probability distribution of the second measurement is determined by the non-selective state $\mathcal{E}(\rho)$ after the first measurement. Especially, we have $q_m(|k\rangle\langle k |) = p_{m}(\mathcal{E}(|k\rangle\langle k |))$.

The classical trace distance between the two probability distributions $\{p_m(\rho)\}$ and $\{q_m(\rho)\}$ is 
\begin{equation}
  D(p_m(\rho),q_m(\rho)) = \frac{1}{2} \sum_m |p_m(\rho) - q_m(\rho)|.
\end{equation}
Based on this classical trace distance, we define the experimental measurement demolition and QND fidelity, respectively, as follows,
\begin{align}\label{eq:ourdestructiveness}
  D_{E} &= \frac{1}{N}\sum_k D(p_m(|k\rangle\langle k |),q_m(|k\rangle\langle k |)), \\
  Q_{E} &= 1 - D_{E}.
\end{align}

For a QND measurement, we have Eq.~(\ref{eq:QNDKraus2}), which is $|k\rangle\langle k | = \mathcal{E}(|k\rangle\langle k |)$ for arbitrary $k$. Then one obtains
\begin{equation}
q_m(|k\rangle\langle k |) = p_{m}(\mathcal{E}(|k\rangle\langle k |)) = p_m(|k\rangle\langle k |).
\end{equation}
In this case, we have $D_{E} = 0$ and $Q_{E} = 1$. To sum up, for a QND measurement, there is $Q_{E} = 1$. Equivalently, if one has $Q_{E} \neq 1$, then the measurement is certainly not a QND measurement.

Besides, in appendix \ref{appendix:QNDKrausQNDNewDefinition}, we have proved that a QND measurement implies the following
\begin{equation}\label{eq:QNDNewDefinition}
  q_{m}(\rho)=p_m(\rho), ~\forall m,\rho.
\end{equation}
Conversely, when $\{E_m\}$ are all diagonal and linearly independent, Eq.~(\ref{eq:QNDNewDefinition}) induces Eq.~(\ref{eq:QNDKraus}) and thus implies a QND measurement.

\subsection{Relationship with theoretical QND fidelity}
For density matrices $\rho$ and $\sigma$, we denote ${p'_m(\rho)} = {\mathrm{Tr}[ F_m \rho]}$ and ${q'_m(\sigma)} = {\mathrm{Tr}[ F_m \sigma]}$, where $F_m$ is a POVM element. The quantum trace distance relates closely to the classical trace distance \cite{nielsen2010quantum},
\begin{equation}
  D(\rho,\sigma) = \max \limits_{\{F_m\}} D(p'_m(\rho),q'_m(\sigma)),
\end{equation}
where the maximization is over all POVMs $\{F_m\}$.
Therefore, for any POVM element $F_m$, we have that \cite{nielsen2010quantum}
\begin{equation}
  D(\rho,\sigma) \geq D(p'_m(\rho),q'_m(\sigma)).
\end{equation}
Substituting $\rho = |k\rangle\langle k |$, $\sigma=\mathcal{E}(|k\rangle\langle k |)$, and $F_m=E_m$ into the above equation, one obtains that
\begin{equation}\label{eq:TraceDistanceEquality}
D(|k\rangle\langle k |,\mathcal{E}(|k\rangle\langle k |)) \geq D(p_m(|k\rangle\langle k |),q_m(|k\rangle\langle k |)),
\end{equation}
Thus it stands that,
\begin{align}
  D_{D} &\geq D_{E}, \\
  Q_{D} &\leq Q_{E}.
\end{align}
Note that a QND measurement induces $Q_{E}=Q_{D}=1$. To conclude, the experimental $Q_{E}$ provides a reachable upper bound of the theoretical one $Q_D$. While $Q_D=1$ certainly implies a QND measurement, $Q_E=1$ is a weaker characterization of QND measurements. To be specific, $Q_{E}=1$ is weaker than $Q_{D}=1$, that is $Q_{D}=1$ induces $Q_{E}=1$, while $Q_{E}=1$ does not necessarily imply $Q_D=1$.

\subsection{Comparison with prior work}
To characterize the back-action evading nature of a QND measurement, L. Pereira \emph{et. al.} introduced a precise quantity based on a self-consistent tomography, to bound the destructiveness
\begin{equation}\label{eq:Destructiveness}
  D_{P} = \frac{1}{2} \max \limits_{\Vert O \Vert = 1}\Vert O-\mathcal{E}^{\dagger}(O) \Vert,
\end{equation}
where $O$ is an arbitrary diagonal observable and satisfies $\Vert O \Vert=1$ \cite{pereira2021complete}. According to Eq.~(\ref{eq:DestructivenessDefination}), a necessary and sufficient condition for a QND measurement is 
\begin{equation}
  D_{P} = 0,
\end{equation}
which is also equivalent to $D_D = 0$, as discussed in Sec.~\ref{sec:notation} and Sec.~\ref{sec:QNDdefinition}.

The advantage of this method based on tomography is that it can completely reconstruct the measurement process. However, its complexity is relatively high. In the above, we denote $N$ as the number of measurement outcomes and the dimension of Hilbert space. For each measurement operator, the tomography complexity is $O(N^4)$ \cite{pereira2021complete}. Since the tomography of $N$ measurement operators is required, the total complexity is $O(N^5)$.

In our proposed experimental method, we calculate the classical trace distance for $N$ computational basis states. For each calculation of the probability distribution distance, $2N$ diagonal elements need to be calculated. Thus the complexity of our proposed method is $O(N^2)$.

In summary, our experimental method is more efficient than that in Ref. \cite{pereira2021complete}. However, unlike the prior work, the method presented in this paper cannot provide complete information about the measurement process. Fortunately, in Sec.~\ref{sec:relationship}, we show that the three key factors, i.e., readout fidelity, projectivity, and QND fidelity can be directly derived with the same experimental scheme.

\section{readout fidelity, projectivity and their relationships with QND fidelity}\label{sec:relationship}
\subsection{Readout fidelity}
The readout fidelity of a measurement is defined as the average probability of obtaining the outcome $k$ when measuring the computation basis state $|k\rangle\langle k |$,
\begin{equation}\label{eq:readoutfidelitydefinition}
F = \frac{1}{N} \sum_k p_{k}(|k\rangle\langle k|) = \frac{1}{N}\sum_k \langle k |E_k|k\rangle,
\end{equation}
where $E_k$ is the POVM element of the measurement. When using the experimental scheme in Fig.~\ref{fig:method2}, only the first measurement and computational basis states are required to obtain the quantity $F$.

It's obvious that $F=1$ if and only if that
\begin{equation}\label{eq:ReadoutFidelity}
\langle k |E_k|k\rangle = 1, \forall k.
\end{equation}
In appendix \ref{appendix:ReadoutFidelity}, we give rigorous proof that Eq.~(\ref{eq:ReadoutFidelity}) stands if and only if that 
\begin{equation}\label{eq:ReadoutFidelity2}
  E_k = |k\rangle\langle k |,~\forall k.
\end{equation}
To conclude, one obtains 
\begin{equation}
  F=1 \iff E_k=|k\rangle\langle k|.
\end{equation}

Note that when $E_k=|k\rangle\langle k|$ is satisfied, for any density matrix $\rho$, one has
\begin{equation}
  p_k(\rho) = \mathrm{Tr}[E_k\rho] = \langle k | \rho |k\rangle,~\forall k, \rho.
\end{equation}
That is to say, when the readout fidelity of a measurement is maximum, one can always obtain the correct probability distribution no matter no matter what the density matrix $\rho$ is.

To conclude, the readout fidelity $F$ evaluates the difference between the POVM element $E_k$ and the projection operator $|k\rangle\langle k |$, and only computation basis states are needed to be taken into consideration.

\subsection{Projectivity}
Using the experimental scheme with two consecutive measurements as shown in Fig.~\ref{fig:method2}, researchers have defined a QND-ness $F_{Q}$ by the average probability of obtaining the same outcome $k$ from the two measurements when measuring the state $\rho = |k\rangle\langle k|$ \cite{lupacscu2007quantum,Picot2010Quantum,Touzard2019Gated,Dassonneville2020Fast}
\begin{align}\label{eq:projectivity}
F_{Q} &= \frac{1}{N} \sum_k p_{k}(|k\rangle\langle k|)p_{k|k}(|k\rangle\langle k|) \\ \nonumber
&=\frac{1}{N}\sum_k \langle k |M_k^{\dagger}E_kM_k|k\rangle.
\end{align}
However, this quantity actually describes the overlap between the actual measurement and an ideal projective one, i.e., projectivity or ideality of the measurement \cite{pereira2021complete}. In this paper, we clarify this definition and name it by projectivity $F_Q$, while in Ref.~\cite{pereira2021complete} that is characterized by the ideality $F_{I}$
\begin{equation}\label{eq:ideality}
  F_{I} = \frac{1}{N}\sum_k |\langle k |M_k|k\rangle|^2.
  \end{equation}

We will prove in the following that these two terms are equivalent by the following facts
\begin{equation}
F_{Q}=1 \iff M_k = m_{k,k} |k\rangle\langle k| \iff F_{I} = 1,
\end{equation}
with $|m_{k,k}|=1$.

Obviously, $M_k = m_{k,k} |k\rangle\langle k|$ induces $F_{Q}=1$. Conversely, when $F_Q=1$ satisfies, one has that
\begin{equation}
1= \langle k |M_k^{\dagger}E_kM_k|k\rangle\leq \langle k |E_k|k\rangle\leq1.
\end{equation}
Thus one obtains $\langle k |M_k^{\dagger}E_kM_k|k\rangle=\langle k |E_k|k\rangle=1$. As in appendix \ref{appendix:ReadoutFidelity}, 
$\langle k |E_k|k\rangle=1$ is valid if and only if $E_k=|k\rangle\langle k|$. Then we have 
\begin{equation}\label{eq:projectivity1}
  |\langle k |M_k|k\rangle|^2 = 1.
  \end{equation}
Employing $\sum_m M_m^{\dagger}M_m = I$ and $\sum_n |n\rangle\langle n| = I$, we have that
\begin{align}\label{eq:projectivity3}
  1 &= \sum_m \langle k | M_m^{\dagger}M_m|k\rangle \\ \nonumber
  &= \sum_{m,n}|\langle n |M_m|k\rangle|^2.
\end{align}
Combing Eq.~(\ref{eq:projectivity1}) and Eq.~(\ref{eq:projectivity3}), if $m\neq k$ or $n \neq k$, one gets that $|\langle n |M_m|k\rangle| = 0$. Finally, one obtains that
\begin{equation}\label{eq:projectivity2}
M_k = m_{k,k} |k\rangle\langle k|,
\end{equation} 
where $|m_{k,k}|=1$. It is also obvious that $M_k = m_{k,k} |k\rangle\langle k|$ implies $F_{I}=1$. Conversely, when one has $F_I=1$, Eq.~(\ref{eq:projectivity1}) is satisfied, which induces Eq.~(\ref{eq:projectivity2}) as proven above.

Since the fact that $|m_{k,k}|=1$, $m_{k,k}$ is a phase and is generally not important in the measurement operator. Without loss of generality, we assume that $m_{k,k}=1$, which induces $M_k =|k\rangle\langle k|$.

In a word, the projectivity shows how close the Kraus operator $M_k$ is to the projection operator $|k\rangle\langle k|$, and only computation basis states are needed to be taken into account.

\subsection{Relationship}
\begin{figure}[tb]
  \centering
  \includegraphics[width=0.48\textwidth]{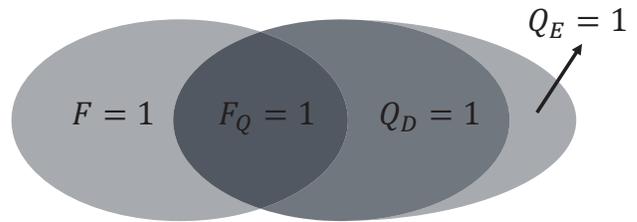}
  \caption{The relationships among readout fidelity $F$, projectivity $F_{Q}$, theoretical QND fidelity $Q_{D}$, and experimental QND fidelity $Q_{D}$. The equation $F=1$ means that POVM elements are projection operators, while $F_{Q}=1$ means that the Kraus operators are projection operators, and the measurement is an ideal projective measurement. The condition $Q_{D}=1$ means that the Kraus operators are diagonal and the measurement is a QND measurement. The experimental QND fidelity $Q_{E}$ is a reachable upper bound of theoretical QND fidelity $Q_{D}$, and thus $Q_{E}=1$ contains $Q_{D}=1$. We conclude that $F_{Q}=1 \iff F=1~\mathrm{and}~Q_D=1$, and $F_{Q}=1 \iff F=1~\mathrm{and}~Q_E=1$ in the main text.}
  \label{fig:relationship}
  \end{figure}
As discussed above, readout fidelity $F$ characterizes how close the POVM elements of the measurement are to projection operators. The maximum value $F=1$ implies that the POVM elements are all projective. Projectivity $F_Q$ quantifies the overlap of the measurement with an ideal projective one. The maximum value $F_Q=1$ implies that the Kraus operators of the measurement are all projective. A maximum value $Q_D=1$ implies a QND measurement, that is the Kraus operators are all diagonal.

Firstly, we will show that the measurement is an ideal projective measurement $F_Q=1$ if and only if both the maximum QND fidelity $Q_D=1$ and readout fidelity $F=1$ are reached, i.e.,
\begin{equation}\label{eq:TheoreticalQNDFideltyandReadoutFidelity}
  F_{Q}=1 \iff F=1~\mathrm{and}~Q_D=1.
  \end{equation}
One can see that $F_Q=1$ implies both $Q_D=1$ and $F=1$. Conversely, a QND measurement with $Q_D=1$ implies that $M_n = \sum_k m_{n,k} |k\rangle\langle k |$. In this case, we have
\begin{equation}
  E_n = \sum_k |m_{n,k}|^2 |k\rangle\langle k |,
\end{equation}
At the same time,  $F=1$ implies that $E_n=|n\rangle\langle n|$. Altogether we have that
\begin{equation}
  M_n = m_{n,n} |n\rangle\langle n |,
\end{equation}
where $|m_{n,n}|=1$. That is to say, the measurement in this case is an ideal projective measurement with $F_Q=1$.

Furthermore, as shown in appendix \ref{appendix:IdealProjectiveMeasurement}, we prove that $F_Q=1$ satisfies if and only if when both the maximum experimental QND fidelity $Q_E=1$ and the maximum readout fidelity $F=1$ are valid, i.e.,
\begin{equation}\label{eq:ExperimentalQNDFideltyandReadoutFidelity}
  F_{Q}=1 \iff F=1~\mathrm{and}~Q_E=1.
  \end{equation}
The equations (\ref{eq:TheoreticalQNDFideltyandReadoutFidelity}) and (\ref{eq:ExperimentalQNDFideltyandReadoutFidelity}) imply that if the readout fidelity reaches maximum value $F=1$, the experimental QND fidelity $Q_E$ and the theoretical QND fidelity $Q_D$ can be used to estimate the overlap of the measurement with an ideal projective measurement.

To conclude, the relationships among readout fidelity, projectivity, theoretical QND fidelity, and experimental QND fidelity of a measurement are shown in Fig.~\ref{fig:relationship}. $Q_D=1$ induces $Q_E=1$. A sufficient and necessary condition for $F_{Q}=1$ is that both $F=1$ and $Q_{D}=1$ are satisfied, which is also equivalent to that both $F=1$ and $Q_{E}=1$ are valid. It is worth noticing that the experimental QND fidelity $Q_E$ not only gives an upper bound of theoretical QND fidelity $Q_D$. Together with readout fidelity $F$, it also can be used to verify whether a measurement is an ideal projective one or not.

\section{theoretical analysis and simulation results }\label{sec:simulation}

\subsection{Theoretical simulation results for dispersive readout}
The experimental QND fidelity is verified by the realistic simulation of the dispersive readout of superconducting qubits. The physical process of superconducting qubit readout is simulated to acquire the three key factors, i.e., readout fidelity, projectivity, and experimental QND fidelity.

\begin{figure}[tb]
  \centering
  \includegraphics[width=8.6cm]{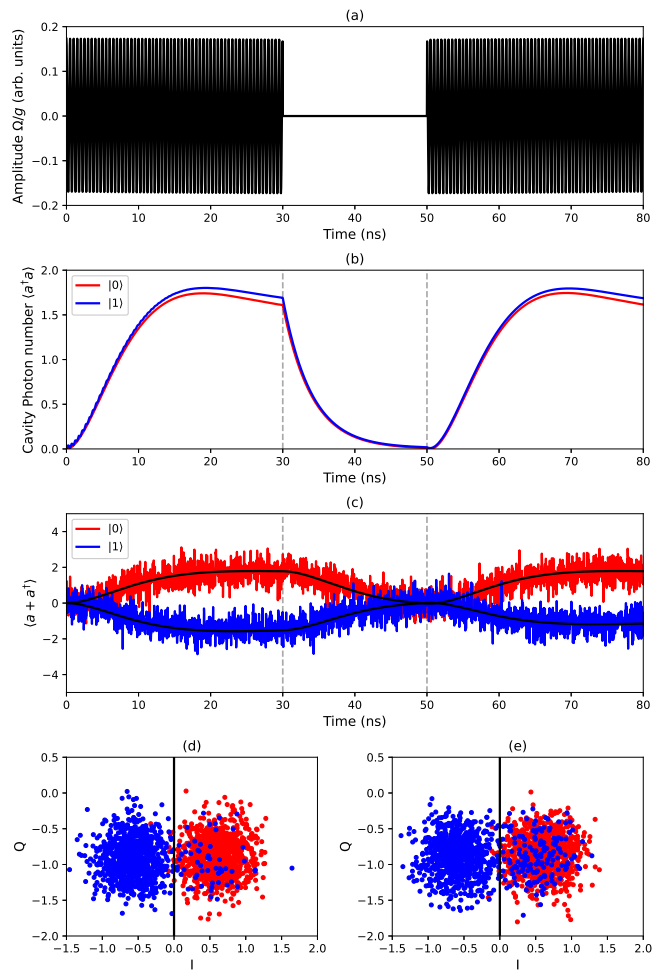}
  \caption{(Color Online) The experimental setup for the theoretical simulations of two consecutive measurements. From 0 to 30 ns, we conduct the first measurement, followed by a period of 20 ns for the reset of the cavity. The second measurement is applied during the final 30 ns. (a) The microwave pulse for readout with time. (b) The photon number in the cavity wth time. (c) Averaged expectation value $\langle a+a^{\dagger} \rangle$ from 1000 trajectories, in which solid black lines show ideal curves based on the Lindblad master equation. (d) The ($I$, $Q$)-records yield a scatter diagram for the first measurement. (d) The ($I$, $Q$)-records yield a scatter diagram for the second measurement. The in-phase $I= \sqrt{\kappa}\int \langle a+a^{\dagger} \rangle dt$ and quadrature $Q= \sqrt{\kappa}\int i\langle a^{\dagger}-a \rangle dt$ components are acquired by a continuos detection. When the qubit is in $|0\rangle$ or $|1\rangle$ state, the sign of the in-phase component $I$ is different, which can be utilized to distinguish the qubit state. The thick black solid lines are separation lines.}
  \label{fig:ExperimentalSetup}
  \end{figure}

In the standard dispersive readout, the qubit couples to a cavity, which is off-resonant from the qubit frequency. Denote the coupling strength as $g$. The detuning between the qubit and the cavity frequency is $\Delta$. The Jaynes-Cummings Hamiltonian of the composite system can be written as
\begin{equation}
  H_{\mathrm{JC}} = \frac{\omega_{q}}{2} \sigma_{z} + \omega_{r}a^{\dagger} a+ g(\sigma_+a+ \sigma_-a^{\dagger}),
\end{equation}
where $\omega_{q}$ and $\omega_{r}$ are the qubit frequency and the cavity frequency, respectively. $\sigma_{z}$ is the Pauli Z matrix. $a$ and $a^{\dagger}$ are the annihilation and creation operators, respectively. Under the dispersive limit, i.e., $\Delta \gg g$, the dispersive Hamiltonian is \cite{Blais2021Circuit}
\begin{equation}
  H_{\mathrm{disp}} \approx \frac{ \omega_{q}}{2} \sigma_{z} +  \omega_{r} a^{\dagger} a+ \chi a^{\dagger} a \sigma_{z},
  \end{equation}
where $\chi = g^2/\Delta$ is the dispersive shift. According to this equation, the frequency shift $\chi$ of the cavity is qubit-dependent. The microwave drive on the cavity can be modeled as \cite{Blais2004Cavity}
\begin{equation}
  H_{d} = \Omega(t)(a^{\dagger}e^{-\omega_{d}t} + a e^{\omega_{d}t}).
\end{equation}
Then the total Hamiltonian can be written as $H=H_{\mathrm{JC}}+H_{d}$. The driving frequency is assumed to be the same with the bare frequency of the cavity, i.e., $\omega_{d}=\omega_{r}$. To be convenient, the Hamiltonian is transformed into the rotating frame, then the Hamiltonian in the interaction picture can be written as
\begin{equation}
  H_{\mathrm{rot}} = g\sigma_+a e^{i\Delta t} + g\sigma_-a^{\dagger}e^{-i\Delta t} + \Omega(t)(a^{\dagger} + a).
\end{equation}
Theoretical simulations are based on this Hamiltonian in this paper. The qubit-dependent frequency shift produces a qubit-dependent displacement on the cavity resonance in the phase space. Finally, the in-phase component $I = \sqrt{\kappa}\int_{0}^{T} \langle a+a^{\dagger} \rangle dt$ obtained by a continuous detection on the cavity varies with the qubit state. Especially, the in-phase component $I$ has the opposite sign when the two-level qubit is in $|0\rangle$ or $|1\rangle$ states. One can discriminate the qubit state based on the sign of the in-phase component.

Our experimental parameters refer to Ref.~\cite{pereira2021complete}. The coupling strength is chosen to be $g/2\pi = 200$ MHz, the decay rate of the cavity is set as $\kappa = 0.2g$, the energy relaxation rate $\gamma$ and the dephasing rate $\gamma_\phi$ of the qubit are $\gamma=\gamma_\phi=10^{-4}g$, and the measurement time is $T = 8/\kappa \approx 30$ ns. The driving amplitude is $\Omega_c = 0.173g$, corresponding to 
$\langle a^{\dagger} a \rangle \approx 1.6$ when $2\chi=\kappa$. The simulated physical process is shown as Fig.~\ref{fig:ExperimentalSetup}, in which we set $2\chi=\kappa$. A microwave pulse is used to drive the cavity, the photon number varies over time. From 0 to 30 ns, we conduct the first measurement, followed by a period of 20 ns for the reset of the cavity. The average photon number of the cavity is as low as 0.01 after the reset. Then the second measurement is applied during the final 30 ns. The in-phase component $I$ has the opposite sign when the qubit is in $|0\rangle$ and $|1\rangle$ states. Figures~\ref{fig:ExperimentalSetup}(d)-(e) presents the $I$-$Q$ scatter diagrams from a continuous heterodyne detection for the twice measurements.

\begin{figure}[tb]
  \centering
  \includegraphics[width=8.6cm]{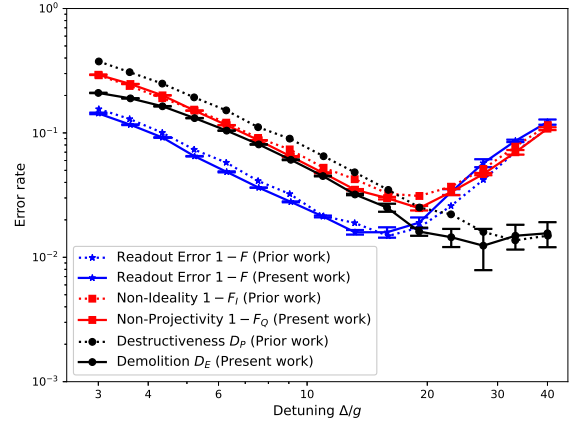}
  \caption{(Color Online) Comparison of theoretical simulation results in this paper to those in the prior work \cite{pereira2021complete}. The horizontal axis is the detuning normalized by the coupling strength $\Delta/g$, while the vertical axis is the error rate. The readout error $1-F$ is from Eq.~(\ref{eq:readoutfidelitydefinition}), the non-ideality $1-F_I$ is from Eq.~(\ref{eq:ideality}), the non-projectivity $1-F_Q$ is from Eq.~(\ref{eq:projectivity}), the demolition $D_E$ is from Eq.~(\ref{eq:ourdestructiveness}), and the destructiveness $D_{P}$ is from Eq.~(\ref{eq:Destructiveness}). The data of dashed lines are from Ref.~\cite{pereira2021complete}, while our simulation results are shown with the solid lines. The simulation is done with 1000 trajectories for each computational basis state. The error bars correspond to the standard deviation obtained with five simulations.}
  \label{fig:simulation}
  \end{figure}

Theoretical simulation of quantum measurement is based on the formalism of a stochastic master equation \cite{gambetta2008quantum,wiseman2009quantum,pereira2021complete}, using the qutip toolkit \cite{johansson2013qutip}. The qubit is assumed as an ideal two-level system. The number of energy levels of the cavity is truncated to 10. The simulation is done with 1000 trajectories for each computational basis state. Figure~\ref{fig:simulation} shows the theoretical simulation results. The horizontal axis is the detuning normalized by the coupling strength, i.e., $\Delta/g$. The vertical axis is the error rate corresponding to the readout error $1-F$, non-ideality $1-F_I$, non-projectivity $1-F_Q$, destructiveness $D_P$, and demolition $D_E$, respectively. The dashed lines are original data from Ref.~\cite{pereira2021complete}, in which the ideality $F_I$ and the destructiveness $D_P$ are defined. The solid lines illustrate our simulation results and the error bars are the standard deviations obtained with five simulations. The readout error and non-projectivity agree well with the readout error and non-ideality in the prior work, respectively. This proves the validity of the theoretical simulation and benchmarks our simulation code.

Most importantly, the demolition $D_E$ proposed in this paper has basically the same tendency as the destructiveness $D_P$ presented in prior work, as shown in Fig.~\ref{fig:simulation}. The minimum value of destructiveness is limited by the qubit decoherence, which breaks the QND condition \cite{pereira2021complete}. In the above discussion, we know that demolition $D_E$ can be obtained efficiently in experiments without using quantum tomography to realize complete physical characterization of the measurement, which is of great significance for practical applications. 

Besides, as shown in Fig.~\ref{fig:simulation}, simulation results present different optimal points for the readout fidelity, projectivity, and QND fidelity, which could be useful for the optimization of measurements from different application requirements. The tendency of the demolition or destructiveness is quite different from the non-projectivity used in Refs.~\cite{lupacscu2007quantum,Picot2010Quantum,Touzard2019Gated,Dassonneville2020Fast} to quantity the QND-ness, which is also studied in Ref.~\cite{pereira2021complete}. In the following, we demonstrate more cases that show the deviation of the projectivity and QND fidelity.

\subsection{Special cases}
The projectivity $F_Q$ is used to evaluate the QND character of a measurement in many works. This is reasonable when the readout fidelity is close to one. However, the QND requirement $[M_m, \sigma_z] = 0$, namely $[H, \sigma_z] = 0$, does not definitely induce the maximum value of projectivity $F_Q=1$. We will show two special cases that one obtains the projectivity $F_Q \neq 1$, even though the measurement is a QND measurement.

For a two-level system, assume that the Kraus operators of a measurement are
\begin{equation}
  M_1 = \left[
    \begin{array}{cc}
      \cos\theta & 0 \\
      0 & \sin\theta
    \end{array}
  \right],~
  M_2 = \left[
    \begin{array}{cc}
      \sin\theta & 0 \\
      0 & \cos\theta
    \end{array}
  \right].
\end{equation}
The Kraus operators are all diagonal, i.e., $[M_m, \sigma_z] = 0$. Thus the measurement is a QND measurement, which induces $Q_D = 1$ and $Q_E = 1$. However, we calculate that $F_Q = \cos^4\theta$ is not equal to one in most cases. The projectivity $F_Q$ is deviated from the characterization of the QND character in this case.

The other case occurs in the readout of superconducting qubits. The qubit is coupled to a readout cavity. Assuming the state of the qubit is $|\varphi\rangle = c_1 |0\rangle + c_2 |1\rangle$, the cavity is initialized in the vacuum state $|0\rangle$. The initial state of the composite system is $|\psi\rangle = |\varphi\rangle \otimes |0\rangle$. If the Hamiltonian of the composite system satisfies $[H, \sigma_z] = 0$, such as the longitudinal readout \cite{Blais2021Circuit}, the steady state would be $|\psi'\rangle = c_1 |\alpha_0\rangle \otimes |0\rangle + c_2 |\alpha_1\rangle\otimes |1\rangle$, where $|\alpha_0\rangle$ and $|\alpha_1\rangle$ refer to the corresponding coherent state of the cavity when the qubit is in the $|0\rangle$ and $|1\rangle$ state, respectively. The Kraus operator of the heterodyne detection of the cavity is
\begin{equation}
  K_\alpha = \frac{1}{\sqrt{\pi}} |\alpha \rangle\langle \alpha |,
\end{equation}
with the corresponding output value being $\alpha$, which is a continuous variable. Finally, we obtain the Kraus operator of the measurement for the qubit as follows,
\begin{equation}
  M_\alpha = \frac{1}{\sqrt{\pi}} \left(\langle\alpha |\alpha_0\rangle |0\rangle\langle 0 | + \langle\alpha |\alpha_1\rangle |1\rangle\langle 1 | \right).
\end{equation}
One sees that the Kraus operators are diagonal and thus the measurement is a QND measurement. Naturally, one obtains $Q_E = Q_D = 1$. However, the Kraus operators are not projective ones, which implies that $F_Q \neq 1$. The separation errors \cite{krantz2019quantum} prevent maximum projectivity.

In summary, the projectivity $F_Q$ is deviated from the characterization of QND measurement in some special cases, while the QND fidelity proposed in this paper quantifies the QND character of the measurement.

\section{conclusion}\label{sec:conclusion}
We quantify the back-action evading nature of a QND measurement in this paper. Only computational basis states are needed to be prepared. The quantum trace distance of the states before and after the measurement is utilized to quantify the QND character. We also propose an experimentally efficient method to directly derive the experimental QND fidelity by using classical trace distance based on an experimental scheme with two consecutive measurements. The experimental QND fidelity not only bounds the theoretical QND fidelity, but also can be used to characterize an ideal projective measurement together with readout fidelity.

The relationships among three key quantifiers of a measurement, i.e., readout fidelity, projectivity, and QND fidelity, are presented in this paper. Besides, these factors can be derived directly through the same experimental scheme using computational basis states.

Theoretical simulation results of superconducting dispersive qubit readout demonstrate the validity of the experimental QND fidelity. The results also present different optimal points for the readout fidelity, projectivity, and QND fidelity, which could be important for the optimization of measurements from different application requirements. We expect our results to find promising applications in characterizing the performance of qubit measurements and be helpful for the diagnosis and improvement of quantum measurement apparatuses.

\begin{acknowledgments}
The authors would like to thank Kun Wang for useful discussions and inspiring comments on the manuscript.
  \end{acknowledgments}

\appendix
\section{EQUIVALENCE AMONG  DIFFERENT DEFINITIONS OF QND MEASUREMENT}\label{appendix:QNDKrausQNDAdjoint}
  \subsection{Equivalence between Eq.~(\ref{eq:QNDAdjoint}) and Eq.~(\ref{eq:QNDKraus})}
  
  \textbf{Proof:} 
  
Eq.~(\ref{eq:QNDKraus}) $\Rightarrow$ Eq.~(\ref{eq:QNDAdjoint}):
  
Since $\{M_n\}$ are all diagonal, and all diagonal matrices are commute, it's straightforward that 
  \begin{align}
    \mathcal{E}^{\dagger}(|k\rangle\langle k |) &= \sum_m M_m^{\dagger} |k\rangle\langle k | M_m \\ \nonumber
    &= \sum_m M_m^{\dagger} M_m |k\rangle\langle k | = |k\rangle\langle k |.
  \end{align}
  
Eq.~(\ref{eq:QNDAdjoint}) $\Rightarrow$ Eq.~(\ref{eq:QNDKraus}):
   
Multiplying $|l\rangle\langle l|$ on both sides of $|k\rangle\langle k | = \mathcal{E}^{\dagger}(|k\rangle\langle k |)$, we have that 
  \begin{equation}
    \delta_{lk} |k\rangle\langle k | = \sum_m |\langle k | M_m |l\rangle|^2 |l\rangle\langle l |.
  \end{equation}
  
Since $|\langle k | M_m |l\rangle|^2\geq 0 $, it must stand that $\langle k | M_m |l\rangle = 0$ for any $k \neq l$ and any $m$. That is to say, 
  \begin{equation}
    M_n = (\sum_k |k\rangle\langle k |)M_m(\sum_l |l\rangle\langle l |) = \sum_k m_{n,k} |k\rangle\langle k |,
  \end{equation}
  where $m_{n,k} = \langle k | M_n |k\rangle$.

 End proof.
  
 \subsection{Equivalence between Eq.~(\ref{eq:QNDKraus}) and Eq.~(\ref{eq:QNDKraus2})}
 Following a similar analysis as above, we can easily derive that Eq.~(\ref{eq:QNDKraus}) $\iff$  Eq.~(\ref{eq:QNDKraus2}).

  \textbf{Proof:} 
  
Eq.~(\ref{eq:QNDKraus}) $\Rightarrow$ Eq.~(\ref{eq:QNDKraus2}):

Since $\{M_n\}$ are all diagonal, and all diagonal matrices are commute, it's straightforward that 
  \begin{align}
    \mathcal{E}(|k\rangle\langle k |) &= \sum_m M_m |k\rangle\langle k | M_m^{\dagger} \\ \nonumber
    &= \sum_m M_m^{\dagger} M_m |k\rangle\langle k | = |k\rangle\langle k |.
  \end{align}

 Eq.~(\ref{eq:QNDKraus2}) $\Rightarrow$ Eq.~(\ref{eq:QNDKraus}):
Multiplying $|l\rangle\langle l|$ on both sides of $|k\rangle\langle k | = \mathcal{E}(|k\rangle\langle k |)$, we have that 
  \begin{equation}
    \delta_{lk} |k\rangle\langle k | = \sum_m |\langle l | M_m |k\rangle|^2 |l\rangle\langle l |.
  \end{equation}
   
Since $|\langle l | M_m |k\rangle|^2 \geq 0$, it must stand that $\langle l | M_m |k\rangle = 0$ for any $k \neq l$ and any $m$. That is to say, 
  \begin{equation}
    M_n = (\sum_k |k\rangle\langle k |)M_m(\sum_l |l\rangle\langle l |) = \sum_k m_{n,k} |k\rangle\langle k |,
  \end{equation}
  where $m_{n,k} = \langle k | M_n |k\rangle$.

End proof. 

\section{RELATIONSHIP BETWEEN Eq.~(\ref{eq:QNDKraus}) AND Eq.~(\ref{eq:QNDNewDefinition})}\label{appendix:QNDKrausQNDNewDefinition}
By definition, $\{p_m(\rho)\}$ and $\{q_m(\rho)\}$ are the first and second measurement probability distributions, respectively, for a density matrix $\rho$. Notice that, 
\begin{equation}
  q_{m}(\rho) = p_{m}(\mathcal{E}(\rho)) = \mathrm{Tr}[E_m \mathcal{E}(\rho)] = \mathrm{Tr}[\rho\mathcal{E}^{\dagger}(E_m)],
\end{equation}
where $p_m(\rho) = \mathrm{Tr}[E_m \rho]$. Therefore, $q_m(\rho) = p_m(\rho)$ satisfies if and only if that, 
\begin{equation}
    E_m = \mathcal{E}^{\dagger}(E_m),~\forall m.
\end{equation}

In the following, we will show that the above equation is equivalent to Eq.~(\ref{eq:QNDKraus}) with assumptions that $\{E_m\}$ are all diagonal and linear independent.

\textbf{Proof}:

Eq.~(\ref{eq:QNDKraus}) $\Rightarrow$ Eq.~(\ref{eq:QNDNewDefinition}):

As in Eq.~(\ref{eq:QNDKraus}), $\{M_n\}$ are all diagonal, and diagonal matrices are all commute. We have the following,
  \begin{equation}
    \mathcal{E}^{\dagger}(E_m) = \sum_n M_n^{\dagger} E_m M_n = \sum_n M_n^{\dagger} M_n E_m = E_m.
  \end{equation}
 with $E_m = M_m^{\dagger}M_m$.

Eq.~(\ref{eq:QNDNewDefinition}) $\Rightarrow$ Eq.~(\ref{eq:QNDKraus}):

Together with assumptions that $\{E_m\}$ are all diagonal and linear independent, it's straightforward that,
  \begin{equation}
    \mathcal{E}^{\dagger}(|k\rangle\langle k |) = \sum_{m} p_{k,m} \mathcal{E}^{\dagger}(E_m) = \sum_{m} p_{k,m} E_m = |k\rangle\langle k |,~\forall k.
  \end{equation}
with $|k\rangle\langle k | = \sum_{m} p_{k,m} E_m$ and $p_{k,m} = \langle k |E_m|k\rangle$.

According to the result in appendix \ref{appendix:QNDKrausQNDAdjoint}, the above equation implies Eq.~(\ref{eq:QNDKraus}).

End proof.

\section{EQUIVALENCE BETWEEN Eq.~(\ref{eq:ReadoutFidelity}) AND Eq.~(\ref{eq:ReadoutFidelity2})}\label{appendix:ReadoutFidelity}
\textbf{Proof}:

Eq.~(\ref{eq:ReadoutFidelity2}) $\Rightarrow$ Eq.~(\ref{eq:ReadoutFidelity}):

Obviously,  $E_k = |k\rangle\langle k |$ implies that,
  \begin{equation}
    \langle k |E_k|k\rangle = 1, \forall k.
  \end{equation}
    
Eq.~(\ref{eq:ReadoutFidelity}) $\Rightarrow$ Eq.~(\ref{eq:ReadoutFidelity2}):

Notice that $\sum_m E_m = I$, we have,
\begin{align}
  1 &= \langle k |\sum_m E_m|k\rangle = \sum_{m\neq k} \langle k |E_m|k\rangle + \langle k |E_k|k\rangle \nonumber \\
  &= 1 + \sum_{m\neq k} \langle k |E_m|k\rangle,
\end{align}
Thus, it must stand that,
\begin{equation}
  \langle k |E_m|k\rangle =\delta_{m,k}.
\end{equation}

Since $\langle k |E_m| k \rangle=\delta_{m,k}$ and $E_m=E_m^{\dagger}$, we have that 
  \begin{equation}
    E_m=\sum_{i,j}e_{i,j}|i\rangle\langle j|,
  \end{equation}
  with $e_{i,i}=\delta_{i,m}$ and $e_{i,j}=\bar{e}_{j,i}$.

Since $E_m=M_m^{\dagger}M_m\geq 0$, it stands that $\langle \psi |E_m| \psi \rangle\geq 0$ for any (unnormalized) vector $| \psi \rangle$.

Taking $| \phi_{k,j} \rangle=|k\rangle + x|j\rangle$ for $j \neq k$, $\langle \phi_{k,j} |E_m |\phi_{k,j}\rangle= \delta_{k,m} + \delta_{j,m} + e_{j,k}\bar{x}+e_{k,j}x =\delta_{k,m} + \delta_{j,m} +2\mathrm{Re}[e_{k,j}x] \geq 0$. For any nonzero $e_{k,j}$, there always exists an $x$ such that the above inner product is less than 0. Thus $e_{k,j}=e_{j,k}=0$ for any $k \neq j$. Together with $e_{i,i}=\delta_{i,m}$, one finally obtains that
\begin{equation}
  e_{i,j}=\delta_{i,j}\delta_{i,m},
\end{equation}
which is equivalent to that
\begin{equation}
  E_m = |m\rangle\langle m |,~\forall m.
\end{equation}

\section{IDEAL PROJECTIVE MEASUREMENT}\label{appendix:IdealProjectiveMeasurement}
In this section, we show that the 
\begin{equation}
  F_{Q}=1 \iff F=1~\mathrm{and}~Q_E=1.
  \end{equation}

At first, it's easy to prove that when $F_Q=1$, it must have that $Q_E=1$ and $F=1$.
On the counterpart, $Q_E=1$ implies that $p_{m}(\mathcal{E}(|k\rangle\langle k |)) = p_m(|k\rangle\langle k |)$. That is to say, 
   \begin{equation}
    \sum_n \langle k |M_n^{\dagger}E_mM_n|k\rangle=\langle k |E_m|k\rangle.
  \end{equation}
At the same time, $F=1$ implies that $E_m=|m\rangle\langle m|$. Substituting it into the above equation, we have the following,
\begin{equation}
\sum_{n}|\langle m |M_n|k\rangle|^2 = \delta_{m,k}.
\end{equation}
Then one obtains $\{M_n\}$ are all diagonal, i.e., $Q_D=1$. Using Eq.~(\ref{eq:TheoreticalQNDFideltyandReadoutFidelity}), one obtains $F_Q=1$.


\bibliography{apssamp}

\end{document}